\documentclass[aip,jmp,reprint]{revtex4-1}
\usepackage{graphicx}
\usepackage{amsmath}
\usepackage{amssymb}
\newtheorem{thm}{Theorem}[section]
\newtheorem{cor}[thm]{Corollary}
\newtheorem{lem}[thm]{Lemma}
\newtheorem{prop}[thm]{Proposition}
\newtheorem{defn}[thm]{Definition}

\numberwithin{equation}{section}

\newcommand{\supp}[1]{\text{supp}(#1)}
\newcommand{\CH}[1]{\text{ConvHull}(#1)}

\begin{document}
\title{The Motion of a Body in Newtonian Theories}
\author{James Owen Weatherall}
\email{weatherj@uci.edu}
\affiliation{Department of Logic and Philosophy of Science, University of California--Irvine, 3151 Social Science Plaza A, Irvine, CA 92697}
\begin{abstract}
A theorem due to Bob Geroch and Pong Soo Jang [``Motion of a Body in General Relativity.'' \emph{Journal of Mathematical Physics} \textbf{16}(1), (1975)] provides the sense in which the geodesic principle has the status of a theorem in General Relativity (GR).  Here we show that a similar theorem holds in the context of geometrized Newtonian gravitation (often called Newton-Cartan theory).  It follows that in Newtonian gravitation, as in GR, inertial motion can be derived from other central principles of the theory.
\end{abstract}
\maketitle

\section{Introduction}\label{sec-introduction}
The geodesic principle in General Relativity (GR) states that free massive test point particles traverse timelike geodesics.  It has long been believed that, given the other central postulates of GR, the geodesic principle can be proved as a theorem.  In our view, though previous attempts \cite{*[{For a sampling, see }] Einstein+etal, *Thomas, *Taub, *Dixon, *Souriau, *[{ and }] Sternberg} were highly suggestive, the sense in which the geodesic principle is a theorem of GR was finally clarified by Bob Geroch and Pong Soo Jang.\cite{Geroch+Jang, *[{ See also }] [{ who prove a version of the Geroch-Jang theorem that permits backreaction of the particle on the metric.}] Ehlers+Geroch}  They proved the following (the statement of which is indebted to \citet[Prop. 2.5.2]{MalamentGR}):
\begin{thm}\label{GJ}
\emph{(\textbf{\citet{Geroch+Jang}})}
Let $(M,g_{ab})$ be a relativistic spacetime, with $M$ orientable.  Let $\gamma:I\rightarrow M$ be a smooth, imbedded curve.  Suppose that given any open subset $O$ of $M$ containing $\gamma[I]$, there exists a smooth symmetric field $T^{ab}$ with the following properties.
\begin{enumerate}
\item \label{sdec} $T^{ab}$ satisfies the \emph{strengthened dominant energy condition}, i.e. given any future-directed timelike covector $\xi_a$ at any point in $M$, $T^{ab}\xi_a\xi_b\geq 0 $ and either $T^{ab}=\mathbf{0}$ or $T^{ab}\xi_a$ is timelike;
\item \label{cons}$T^{ab}$ satisfies the \emph{conservation condition}, i.e. $\nabla_a T^{ab}=\mathbf{0}$;
\item \label{inside}$\supp{T^{ab}}\subset O$; and
\item \label{non-vanishing}there is at least one point in $O$ at which $T^{ab}\neq \mathbf{0}$.
\end{enumerate}
Then $\gamma$ is a timelike curve that can be reparametrized as a geodesic.
\end{thm}
The interpretation of the Geroch-Jang theorem can be put as follows: if $\gamma$ is a smooth curve about which it is possible to construct an arbitrarily small matter field satisfying the conservation and strict dominant energy conditions, then $\gamma$ can be reparametrized as a timelike geodesic.  More roughly, the only curves about which matter can propagate are timelike geodesics.

The Geroch-Jang approach has many virtues that previous attempts lacked:\cite{Einstein+etal} (1) Geroch and Jang do not make any specific assumptions about the kinds of matter fields that might compose the free massive test point particle (i.e. they do not need to assume it is a perfect fluid or a dust, etc.), aside from general assumptions that \emph{any} body in GR would be expected to satisfy; (2) Geroch and Jang are able to show that a free massive test point particle traverses a curve \emph{within spacetime}, as opposed to a ``line singularity''; and (3) Geroch and Jang do not need to make simplifying assumptions regarding the mass multi-pole structure of their test objects.

In so-called ``geometrized Newtonian gravitation'' (sometimes, ``Newton-Cartan theory''), a reformulation of Newtonian gravitation first developed in the 1920s by \'{E}lie Cartan\cite{Cartan1, *Cartan2} and Kurt Friedrichs,\cite{Friedrichs} with substantial later contributions by Ehlers,\cite{Ehlers} K\"unzle,\cite{Kunzle} and Trautman,\cite{Trautman} (see \citet[Ch. 4]{MalamentGR} for an extensive list of references) the motion of a free massive test point particle is again governed by a geodesic principle.  But thus far, little attention has been paid to the question of whether here, too, the geodesic principle has the status of a theorem.\cite{*[{The exception is }] [{, who use variational methods to derive the equations of motion of a particle in geometrized Newtonian gravitation.  The present result should be understood to bear the same relationship to Duval and Kunzle's work that the Geroch-Jang theorem bears to attempts to derive the geodesic principle using variational methods in GR.  They represent different routes to (almost) the same place.  It is our view that for some purposes, the Geroch-Jang method is considerably more perspicuous.}] Duval+Kunzle}  The central result of the present paper (Theorem \ref{W}) is that a direct parallel to the Geroch-Jang theorem does hold in geometrized Newtonian gravitation.\footnote{\label{Harvey}At least, the Geroch-Jang theorem and Theorem \ref{W} of this paper are directly parallel mathematically.  There is a second kind of question that one might ask, concerning the interpretations of the two theorems in the contexts of their respective spacetime theories. For instance, one might wonder if the conservation condition is as natural an assumption in geometrized Newtonian gravitation as in GR.  We do not address such questions here, but will return to them in future work.}  It is worth noting that in the course of proving the geodesic principle as a theorem of geometrized Newtonian gravitation, we prove a lemma that can be understood as a proof of Newton's first law (appropriately reformulated in covariant, four dimensional language) in non-geometrized Newtonian gravitation.  Thus we show that the principles governing inertial motion in both standard Newtonian theory and geometrized Newtonian gravitation are dependent on the other principles of the theory, just as in GR.

The remainder of the paper will proceed as follows. In section \ref{sec-prelimMath}, we will give some preliminary definitions.  The main results of the paper will be presented in section \ref{sec-theorem}, followed by some concluding remarks in section \ref{sec-conclusion}.  A brief review of geometrized Newtonian gravitation is given in appendix \ref{sec-N-C_theory}; appendix \ref{sec-integration} describes some elementary results concerning integration in classical spacetimes that, to our knowledge, have not been considered before and so are offered for completeness. Finally, appendix \ref{sec-proofs} contains proofs of some of the preliminary propositions and lemmas given in sections \ref{sec-prelimMath} and \ref{sec-theorem}.


\section{Some preliminary definitions}\label{sec-prelimMath}

Throughout this section, let $(M,t_a,h^{ab},\nabla)$ be a classical spacetime.  We assume that $\nabla$ is a flat derivative operator and that $M$ is oriented and simply connected.  Let $T^{ab}$ be a smooth symmetric tensor field on $M$ satisfing three conditions: (1) the mass condition, (2) the conservation condition, and (3) given any spacelike hypersurface $\Sigma\subset M$, $\supp{T^{ab}}\cap\Sigma$ is bounded.  We also take for granted some facts and conventions about orientation, volume elements, and hypersurfaces that are described in appendix \ref{sec-integration}.  Finally, we will explicitly indicate that various fields are smooth in the statements of lemmas and theorems, but throughout the supporting discussion, we will at times take for granted than any object that is a candidate for smoothness is indeed smooth.

For any manifold $A$, we will denote the space of all smooth tensor fields on $A$ by $\mathfrak{T}(A)$; the space of smooth contravariant fields on $A$ will be $\mathfrak{T}^{\bullet}(A)$ and the smooth covariant fields on $A$ will be $\mathfrak{T}_{\bullet}(A)$.  Suppose then that $\Sigma\subset M$ is an imbedded submanifold of $M$.  (Note that we will \emph{always} assume that submanifolds are connected.)  The map $\overset{\Sigma}{\imath}:\Sigma\rightarrow M$ will be assumed to represent the imbedding map (i.e. the identity map); the corresponding pull-back map $\overset{\Sigma}{\imath}\,^*:\mathfrak{T}_{\bullet}(M)\rightarrow\mathfrak{T}_{\bullet}(\Sigma)$ represents the restriction of a covariant tensor field on $M$ to a covariant tensor field on $\Sigma$.    Throughout this section and the next, we will write that a given spacelike hypersurface \emph{slices} the support (or the convex hull, etc.) of $T^{ab}$.  This assertion can be spelled out in a number of ways; one that is adequate for current purposes is as follows.  Let $\Sigma\subset M$ be a spacelike hypersurface of $M$.  We will say that $\Sigma$ slices the support (say) of $T^{ab}$ if and only if $\supp{T^{ab}}\cap\Sigma\neq\emptyset$ and for any spacelike hypersurface $\tilde{\Sigma}$ such that $\Sigma\subseteq\tilde{\Sigma}$, $\supp{T^{ab}}\cap\Sigma=\supp{T^{ab}}\cap\tilde{\Sigma}$.  The idea is that there is at least one point $q\in\supp{T^{ab}}$ that is also in $\Sigma$, and moreover, any points in $\supp{T^{ab}}$ that are spacelike related to $q$ are also in $\Sigma$.

We can now establish some basic facts that will be useful in the next section.
\begin{defn}
Given any oriented hypersurface $\Sigma\subset M$, we define the \emph{momentum flux} through $\Sigma$ to be $P^a(\Sigma)=\int_{\Sigma} T^{ab}t_b \overset{\Sigma}{\epsilon}_{cde}=\int_{\Sigma}p^a\overset{\Sigma}{\epsilon}_{cde}$.
\end{defn}
\begin{prop}\label{consMom}
Let $\Sigma_1$, $\Sigma_2$ be any two future-directed spacelike hypersurfaces slicing the support of $T^{ab}$.  Then $P^a(\Sigma_1)=P^a(\Sigma_2)$.
\end{prop}
This proposition follows simply from Stokes' theorem.  Since we will refer to details of the argument in section \ref{sec-theorem}, a proof is given in appendix \ref{sec-proofs}.

If $T^{ab}$ is understood as the Newtonian mass-momentum tensor, Prop. \ref{consMom} is a statement of conservation of momentum.  To see why, note that if $\Sigma_1$ and $\Sigma_2$ are spacelike hypersurfaces slicing the support of $T^{ab}$, then the momentum flux is the same through both of them.  Prop. \ref{consMom} suggests the following definition.

\begin{defn}
Let $\Sigma\subset M$ be any spacelike hypersurface slicing the support of $T^{ab}$.  Then the \emph{total momentum} of the system can be defined pointwise as follows.  At any point $p\in M$, $(P^a)_{|p}=P^a(\Sigma)$.  By Prop. \ref{consMom}, $P^a$ is independent of the choice of surface.
\end{defn}

\begin{prop}
The covariant derivative of $P^a$ is given by $\nabla_nP^a=0$.\end{prop}
This is obvious, though a proof can be given along the lines of the proof of Prop. \ref{delAngMom} given in appendix \ref{sec-proofs}.  Note that $P^a$ is timelike, as $P^{a}t_a=\int_{\Sigma} T^{ab}t_at_b\overset{\Sigma}{\epsilon}_{cde}>0$, and so $P^a$ is a constant timelike vector field relative to $\nabla$.  Thus its integral curves are geodesics.  It is convenient to work with a normalized vector field, $V^a$, given by $V^a=P^a/(P^nt_n)$, whose integral curves are also geodesics.  In what follows, let $\Gamma$ be the set of maximal integral curves of $V^a$.

Since $\nabla$ is flat, we can define a class of vector fields, $\{\overset{p}{\chi}\,^a|p\in M\}\subset\mathfrak{T}^{\bullet}$, satisfying the following properties: for any $p\in M$, $(\overset{p}{\chi}\,^a)_{|p}=\mathbf{0}$ and $\nabla_a\overset{p}{\chi}\,^b=\delta_{a}{}^b$.\cite{MalamentGR}  These can be thought of as fields of ``position vectors'' centered at a specified point.  At each point $q$, $(\overset{p}{\chi}\,^a)_{|q}$ gives the vector ``from $p$ to $q$'' in the tangent space at $q$. These position fields allow us to define angular momentum flux in the geometrized context.

\begin{defn}
Given any point $p\in M$ and any oriented hypersurface $\Sigma\subset M$, we define the \emph{angular momentum flux} through $\Sigma$ relative to $p$ to be $J^{ab}(\Sigma,p)=\int_{\Sigma}\overset{p}{\chi}\,^{[a}T^{b]c}t_c\overset{\Sigma}{\epsilon}_{def}$.
\end{defn}

\begin{prop}\label{consAngMom}
Let $\Sigma_1$, $\Sigma_2$ be any two future-directed spacelike hypersurfaces slicing the support of $T_{ab}$ and let $p\in M$.  Then $J^{ab}(\Sigma_1,p)=J^{ab}(\Sigma_2,p)$.
\end{prop}
We omit the proof of this claim, as it follows by identical reasoning as the proof of Prop. \ref{consMom}.  Prop. \ref{consAngMom} is analogous to Prop. \ref{consMom} and can similarly be interpreted as a statement of the conservation of angular momentum about any given point.  It justifies a definition analogous to that of $P^a$.

\begin{defn}
Let $\Sigma\subset M$ be any spacelike hypersurface slicing the support of $T^{ab}$.  Then the \emph{total angular momentum}, $J^{ab}$, can be defined pointwise in the following way.  At any point $p\in M$, $(J^{ab})_{|p}=J^{ab}(\Sigma,p)$.  By Prop. \ref{consAngMom}, $J^{ab}$ at any point is independent of the choice of $\Sigma$.\end{defn}

\begin{prop}\label{delAngMom}
The covariant derivative of $J^{ab}$ is given by $\nabla_a J^{bc}=-\delta_a{}^{[b}P^{c]}$.
\end{prop}
A proof of this proposition is given in appendix \ref{sec-proofs}.

Now suppose additionally that $(M,\nabla)$ is geodesically complete.  We can use the concepts already defined to describe the center of mass of $T^{ab}$.
\begin{defn}
A set $A\subseteq M$ is \emph{spatially convex} if and only if for all $p,q\in A$ for which there is a spacelike geodesic segment $\gamma:I\rightarrow M$ with endpoints $p$ and $q$, $\gamma[I]\subseteq A$.  For any tensor field $X^{a_1\cdots}_{b_1\cdots}$, let $X=\{\tilde{X}|\tilde{X}\text{ is spatially convex and }\supp{X^{a_1\cdots}_{b_1\cdots}}\subseteq \tilde{X}\}$.  Then the \emph{spatial convex hull} of $X^{a_1\cdots}_{b_1\cdots}$, denoted $\CH{X^{a_1\cdots}_{b_1\cdots}}$, is given by $\CH{X^{a_1\cdots}_{b_1\cdots}}=\bigcap X$. \end{defn}
At times, we will drop the ``spatial,'' but we will always mean the spatial convex hull.

\begin{prop}
\label{uniqueCoM}
Let $\Sigma$ be a spacelike hypersurface slicing the spatial convex hull of $T^{ab}$.  There exists a unique point $q\in \Sigma$ such that $(J^{ab}t_b)_{|q}=\mathbf{0}$.  Moreover, $q\in\CH{T^{ab}}$.
\end{prop}
A proof of this proposition is given in appendix \ref{sec-proofs}.  Prop. \ref{uniqueCoM} allows us to speak of a single center of mass at a given time.

\begin{defn}
Given a spacelike hypersurface $\Sigma$ slicing the spatial convex hull of $T^{ab}$, we will call the unique $q\in \Sigma$ for which $(J^{ab}t_b)_{|q}=\mathbf{0}$ the \emph{center of mass} of $T^{ab}$ in $\Sigma$. \end{defn}

Note finally that since $q\in\CH{T^{ab}}$, we have a sense in which the center of mass is \emph{inside} the worldtube of $T^{ab}$.

\section{A Newtonian geodesic principle}\label{sec-theorem}

We can now consider the motion of a particle in geometrized Newtonian theory.  First, we require several lemmas.  Proofs of the second and third are given in appendix \ref{sec-proofs}; the first is left to the reader.

\begin{lem}\label{flatSpaceGeo}
Let $(M,t_a,h^{ab},\nabla)$ be a classical spacetime, and suppose that $M$ is oriented and simply connected and that $(M,\nabla)$ is geodesically complete. Assume that $\nabla$ is flat.  Let $T^{ab}$ be a smooth symmetric tensor field on $M$ satisfying: (1) the \emph{mass condition}, (2) the \emph{conservation condition}, and (3) given any spacelike hypersurface $\Sigma\subset M$, $\supp{T^{ab}}\cap\Sigma$ is bounded.   Let $G\subset M$ be the collection of center of mass points of $T^{ab}$.  Then there is a smooth curve $(\gamma:I\rightarrow M)\in \Gamma$ (recall that $\Gamma$ is the set of maximal integral curves of $V^a$) such that $G=\gamma[I]$.\end{lem}

It follows immediately that in flat, simply connected, geodesically complete classical spacetimes, the path traced out by the center of mass of $T^{ab}$ can always be reparameterized as a geodesic (so long as $T^{ab}$ is conserved).  In other words, Lemma \ref{flatSpaceGeo} gives us a statement of Newton's first law, as a consequence of the mass condition, the conservation condition, and a condition on the boundedness of the body represented by $T^{ab}$.

The second lemma is more complicated and involves a general classical spacetime.
\begin{lem}\label{flatOp}
Let $(M,t_a,h^{ab},\nabla)$ be a classical spacetime and suppose $M$ is simply connected.  Moreover, suppose that $R^{ab}{}_{cd}=\mathbf{0}$. Let $\gamma:I\rightarrow M$ be a smooth timelike curve.  Then there exists a flat derivative operator on $M$, $\overset{f}{\nabla}$,  that (1) is compatible with $h^{ab}$ and $t_a$ and (2) agrees with $\nabla$ on $\gamma$.\end{lem}
It is important to note that Lemma \ref{flatOp} only provides a flat derivative operator that agrees with $\nabla$ on \emph{timelike} curves.  The argument in appendix \ref{sec-proofs} fails for curves that intersect the same spacelike hypersurface more than once.  This will complicate the proof of the result in the present paper, relative to the Geroch-Jang theorem, but it is not fatal, in large part because of the following result.

\begin{lem}\label{consMass}
Let $(M,t_a,h^{ab},\nabla)$ be an arbitrary classical spacetime, and suppose that $M$ is oriented and simply connected.  Suppose also that $R^{abcd}=\mathbf{0}$.  Let $T^{ab}$ be a smooth symmetric tensor field on $M$ satisfying: (1) the \emph{mass condition}, (2) the \emph{conservation condition}, and (3) given any spacelike hypersurface $\Sigma\subset M$, $\supp{T^{ab}}\cap\Sigma$ is bounded.  Suppose that $\Sigma_1$ and $\Sigma_2$ are spacelike hypersurfaces slicing the support of $T^{ab}$.  Finally, let $\overset{f}{\nabla}$ be \emph{any} flat derivative operator on $M$ that is compatible with the spatial and temporal metrics.  Then $t_aP^a(\Sigma_1)=t_aP^a(\Sigma_2)$, where $P^a(\Sigma_i)$ is defined relative to $\overset{f}{\nabla}$.
\end{lem}

It is now possible to state the general theorem concerning the Newtonian geodesic principle.
\begin{thm}
\label{W}
Let $(M,t_a,h^{ab},\nabla)$ be a classical spacetime, and suppose that $M$ is oriented and simply connected.  Suppose also that $R^{ab}{}_{cd}=\mathbf{0}$.  Let $\gamma:I\rightarrow M$ be a smooth imbedded curve.  Suppose that given any open subset $O$ of $M$ containing $\gamma[I]$, there exists a smooth symmetric field $T^{ab}\in\mathfrak{T}^{\bullet}(M)$ with the following properties.
\begin{enumerate}
\item\label{mass} $T^{ab}$ satisfies the mass condition, i.e. whenever $T^{ab}\neq \mathbf{0}$, $T^{ab}t_at_b>0$;
\item\label{cons2} $T^{ab}$ satisfies the conservation condition, i.e. $\nabla_a T^{ab}=\mathbf{0}$;
\item\label{inside2} $\supp{T^{ab}}\subset O$; and
\item\label{non-vanishing2} there is at least one point in $O$ at which $T^{ab}\neq \mathbf{0}$.
\end{enumerate}
Then $\gamma$ is a timelike curve that can be reparametrized as a geodesic.
\end{thm}
\textbf{Proof.} We will consider three cases.

\textbf{Case 1:} First, suppose that $\gamma$ is (everywhere) timelike.  Let $O$ be an open subset of $M$ containing $\gamma[I]$ and let $T^{ab}$ be a field meeting the requirements of the statement of the theorem.  Since $M$ is always locally geodesically complete, we can freely choose $O$ so that there always exist geodesically complete spacelike hypersurfaces slicing the support of $T^{ab}$.  By Lemma \ref{flatOp}, there exists a flat derivative operator on $M$, $\overset{f}{\nabla}$, that is consistent with $t_a$ and $h^{ab}$, and which agrees with $\nabla$ on $\gamma$.  For each spacelike hypersurface slicing the support of $T^{ab}$, $\Sigma$, it is possible to define $P^a(\Sigma)$ and $J^{ab}(\Sigma)$ (again, we can limit attention to geodesically complete hypersurfaces if necessary).  These fields are defined relative to $\overset{f}{\nabla}$ in the sense that the parallel transport necessary to make sense of such integrals is performed relative to $\overset{f}{\nabla}$.  Note that $P^a(\Sigma)$ and $J^{ab}(\Sigma)$ are globally defined fields; however, since $T^{ab}$ is not necessarily conserved relative to $\overset{f}{\nabla}$, Props. \ref{consMom} and \ref{consAngMom} no longer hold and the fields are dependent on the choice of $\Sigma$.  However, since each $\Sigma$ is geodesically complete, Prop. \ref{uniqueCoM} still holds for each $\Sigma$; likewise Lemma \ref{flatSpaceGeo} continues to hold for each of the $P^a(\Sigma)$ and $J^{ab}(\Sigma)$ fields individually (at least within a neighborhood of the unique center of mass point associated with $\Sigma$), relative to $\overset{f}{\nabla}$.  Thus for each $\Sigma$, there is a geodesic $\overset{\Sigma}{\gamma}$ (relative to $\overset{f}{\nabla}$) that passes through the spatial convex hull of $T^{ab}$ (relative to $\overset{f}{\nabla}$).

As has already been mentioned, $T^{ab}$ is not necessarily conserved relative to $\overset{f}{\nabla}$.  However, $\overset{f}{\nabla}_a T^{ab}=(\overset{f}{\nabla}_a-\nabla_a)T^{ab}$ is given by a smooth field that vanishes on $\gamma$, since by construction the two operators agree there.  Thus, for any constant scalar field $\alpha>0$, one can make $|\overset{f}{\nabla}_aT^{ab}t_b|<\alpha$ everywhere by shrinking the support of $T^{ab}$ (which is always possible because a suitable $T^{ab}$ exists for \emph{any} neighborhood of $\gamma$).

Let $\Sigma_1$ and $\Sigma_2$ be any two appropriate spacelike hypersurface slicing the support of $T^{ab}$ and consider the fields $J^{ab}(\Sigma_1)t_a$ and $J^{ab}(\Sigma_2)t_b$.  The curves $\overset{\Sigma_1}{\gamma}$ and $\overset{\Sigma_2}{\gamma}$ consist of the points at which $J^{ab}(\Sigma_1)t_a$ and $J^{ab}(\Sigma_2)t_a$ vanish, respectively.  Now let $\Sigma$ be some other appropriate spacelike hypersurface slicing the support of $T^{ab}$, and let $p\in\Sigma$. The field $J^{ab}(\Sigma_1)t_a$ (for instance) at $p$ can be interpreted as the vector pointing from $p$ to $o$, where $o$ is the point at which $\overset{\Sigma_1}{\gamma}$ intersects $\Sigma$.  Note that this interpretation makes sense because (1) $\Sigma$ is always a flat space with Euclidean affine structure and (2) $J^{ab}t_a$ is always spacelike (as can be seen immediately by the symmetry properties of $J^{ab}$).  This means that at any $p$ in an appropriate $\Sigma$, the vector $(J^{ab}(\Sigma_1)-J^{ab}(\Sigma_2))t_a$ represents the vector from $p$ to $o$, minus the vector from $p$ to $o'$ (where $o'$ is the point at which $\overset{\Sigma_2}{\gamma}$ intersects $\Sigma$), which is just the vector from $o'$ to $o$.  Note that this difference is independent of $p$, but dependent on the spacelike hypersurface containing $p$.  So we can define a (spacelike) vector field $\eta^a=(J^{ab}(\Sigma_1)-J^{ab}(\Sigma_2))t_b$ whose spatial length at any point $p$ in a spacelike hypersurface slicing the support of $T^{ab}$ represents the distance between the points at which $\overset{\Sigma_1}{\gamma}$ and $\overset{\Sigma_2}{\gamma}$ intersect that spacelike hypersurface.

Our goal will be to show that the spatial length of $\eta^a$ can be made arbitrarily small everywhere.  To see this, note that since $\eta^a$ is always spacelike, there exists a vector $\beta_a$ such that $\eta^a=h^{ab}\beta_b$.  The spatial length of $\eta^a$ is then given by $(h^{ab}\beta_a\beta_b)^{1/2}$.  Pick an arbitrary point $p\in M$ and consider $h^{ab}\beta_a\beta_b=\beta_a\eta^a$ at $p$.  By definition of the terms involved, this last expression can be written in terms of a constant basis $\overset{1}{\sigma}_a,\ldots,\overset{4}{\sigma}_a$ (relative to $\overset{f}{\nabla}$), so that \begin{align}\label{W-Eq1} h^{ab}\beta_a\beta_b&=\sum_{i=1}^4\overset{i}{\beta}\left(\int_{\Sigma_1}\overset{p}{\chi}{}^{[a}T^{b]c}\overset{i}{\sigma}_at_bt_c\overset{\Sigma_1}{\epsilon}_{def}\right.\notag\\
&\;\;\left.- \int_{\Sigma_2}\overset{p}{\chi}{}^{[a}T^{b]c}\overset{i}{\sigma}_at_bt_c\overset{\Sigma_2}{\epsilon}_{def}\right).
\end{align}
By the Stokes' theorem reasoning in the proof of Prop. \ref{consMom}, we can construct a submanifold $S$ with $\Sigma_1$ and $\Sigma_2$ forming partial boundaries, such that,
\begin{align}\label{W-Eq2}
h^{ab}\beta_a\beta_b&=\sum_{i=1}^4\overset{i}{\beta}\int_{S}\overset{f}{\nabla}_{[n}\overset{p}{\chi}{}^{[a}T^{b]c}\overset{i}{\sigma}_{|a}t_b\overset{S}{\epsilon}_{c|def]}.
\end{align}
Again by the reasoning of the proof of Prop. \ref{consMom}, we can show that $\overset{f}{\nabla}_c(\overset{p}{\chi}{}^{[a}T^{b]c}\overset{i}{\sigma}_{a}t_b)= \overset{p}{\chi}{}^{[a}(\overset{f}{\nabla}_cT^{b]c})\overset{i}{\sigma}_{a}t_b$.  This final expression, meanwhile, represents a scalar field that can be made as small as one likes by shrinking the support of $T^{ab}$.  It follows that the righthand side of Eq. \eqref{W-Eq2} can be made arbitrarily small.  And so, for any positive scalar field $\alpha$, one can choose $O$ so that $h^{ab}\beta_a\beta_b<\alpha$.

It follows that for any two appropriate spacelike hypersurfaces $\Sigma_1$ and $\Sigma_2$, the geodesics $\overset{\Sigma_1}{\gamma}$ and $\overset{\Sigma_2}{\gamma}$ can be made arbitrarily close to one another in the sense that, given any two appropriate spacelike hypersurfaces slicing the support of $T^{ab}$, $\Sigma_1$ and $\Sigma_2$, and any open set $A$ containing $\overset{\Sigma_1}{\gamma}[I]$, we can choose $T^{ab}$ so that $\overset{\Sigma_2}{\gamma}[I]\subset A$ as well.  Moreover, for each $\Sigma$, $\overset{\Sigma}{\gamma}$ passes through the intersection of the spatial convex hull (relative to $\overset{f}{\nabla}$) of $T^{ab}$ and $\Sigma$, and so we can conclude that the image of the original curve, $\gamma[I]$, is arbitrarily close to a geodesic (relative to $\overset{f}{\nabla}$), in the same sense.  This last result is only possible if $\gamma$ can itself be reparameterized as a geodesic (relative to $\overset{f}{\nabla}$).  Finally, since $\overset{f}{\nabla}$ agrees with $\nabla$ on $\gamma$, then $\gamma$ must be a geodesic relative to $\nabla$ as well, up to reparameterization.

\textbf{Case 2:} Now suppose $\gamma$ is (everywhere) spacelike.  We claim that there exist open sets containing $\gamma[I]$ for which there does not exist a smooth symmetric field $T^{ab}\in\mathfrak{T}^{\bullet}(M)$ satisfying conditions \ref{mass}-\ref{non-vanishing2}.  Suppose that for any open set containing $\gamma[I]$, such a field did exist. We know that there always exists a flat derivative operator on $M$, so let $\overset{f}{\nabla}$ be any such flat derivative operator.   Since $\gamma$ is everywhere spacelike, there must be some spacelike hypersurface $\Sigma$ such that $\gamma[I]\subseteq\Sigma$.

First, suppose that $\Sigma$ can be chosen to be bounded.  Then we can also freely choose a neighborhood $O$ of $\gamma$ which is also bounded.  Since $M$ is simply connected, it admits a global time function, $t:M\rightarrow\mathbb{R}$, which is unique up to an additive constant.  We can choose $O$ so that there is some value $t'$ of the time function with the following property: if $\Sigma'$ is a spacelike hypersurface whose time value is $t'$, $\Sigma'$ satisfies $\Sigma'\cap O=\emptyset$.    It follows that $T^{ab}$ vanishes on $\Sigma'$, and thus that $P^a(\Sigma')=\mathbf{0}$ (where the integrals are performed relative to the arbitrary flat derivative operator $\overset{f}{\nabla}$).  Thus $P^a(\Sigma')t_a=0$.  Meanwhile, by the mass condition, we know that $P^a(\Sigma)t_a>0$.  Now we can use a slightly modified\footnote{Modified because by our definition, $\Sigma'$ does not slice the support of $T^{ab}$, since $\Sigma'\cap\supp{T^{ab}}=\emptyset$.  But in this special case the argument still goes through.} version of the argument of Lemma \ref{consMass}.  Since $O$ is bounded, we can freely choose some third (timelike) hypersurface $\Sigma''$ (adjusting our choices of $O$ and $\Sigma$ if necessary) s.t. $\Sigma''\cap O=\emptyset$, and such that $\Sigma\cup\Sigma'\cup\Sigma''$ forms the boundary of a four dimensional submanifold of $M$, $S$ (where we reverse the orientation of, say, $\Sigma'$ so that $S$ is outwardly oriented).  We can thus apply the Stokes' theorem argument given in the proofs of Prop. \ref{consMom} and Lemma \ref{consMass} to show that $P^a(\Sigma)t_a=P^a(\Sigma')t_a$, which is a contradiction.

\begin{figure}\centering
\includegraphics[width=.9\columnwidth]{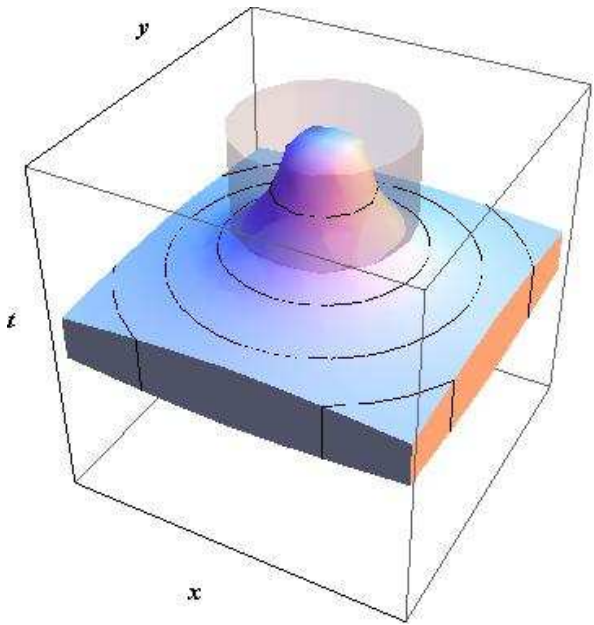}
\caption{\label{spatialInfinity}(Color online.) An example in three dimensions of an open set $O$ whose ``temporal height'' goes to zero at spatial infinity, and which contains a spacelike hypersurface.  (See Case 2 in the text.)}
\end{figure}

Now suppose that $\Sigma$ cannot be chosen to be bounded.  For simplicity, we will assume that $\Sigma$ can be chosen so that it extends to spatial infinity in all directions. (We are ignoring the case where $\Sigma$ is unbounded, but not necessarily in all directions.  The argument given here is intended to be representative: it can be extended to include these more complicated cases by, for instance, choosing $O$ so that the temporal height of its closure would vanish at any boundary of $\Sigma$.) Choose $O$ so that it has the following property: in the limit of spatial infinity, the ``temporal height'' of $O$ goes to zero (see Fig. \ref{spatialInfinity}).  Here is one way (of many) to make this idea precise.  Without loss of generality, choose the time function $t$ so that for any $s\in I$, $t(\gamma(s))=0$.  Let $\varpi$ be any (fixed) timelike geodesic passing through $\Sigma$.  Then given any point $p$ in a spacelike hypersurface intersecting $\varpi$, we can define a distance function $d:M\rightarrow\mathbb{R}$ relative to $\varpi$ as the (spatial) distance from $\varpi$ to $p$.   We can then define an open set $O=\{p\in M |\text{ } |t(p)|<a\text{ and }|d(p)t(p)|<1\}$, for some constant real number $a$ chosen so that $\varpi$ intersects all of the simultaneity slices of $M$ with time values from $-a$ to $a$.  Note first that $\Sigma\subset O$, so $\gamma[I]\subset O$.  Moreover, for any $p\in O-\Sigma$, there exists a spacelike hypersurface $\Sigma'$ for which $p\in\Sigma'$ and $\Sigma'$ slices $O$ (since the restriction of $O$ to any spacelike hypersurface except $\Sigma$ is bounded by construction).

From here the argument is similar to the bounded case. For any given $a$, there exist spacelike hypersurfaces $\Sigma_{\pm}$ such that for any $p\in\Sigma_{+}$, $t(p) > a$, and for any $p\in\Sigma_{-}$, $t(p) < -a$.  These are necessarily such that $\Sigma_{\pm}\cap O=\emptyset$.  It follows that $T^{ab}$ vanishes on $\Sigma_{\pm}$, and thus that $P^a(\Sigma_{\pm})=\mathbf{0}$ (where the integrals are performed relative to the arbitrary flat derivative operator $\overset{f}{\nabla}$).  Thus $P^a(\Sigma_{\pm})t_a=0$.  Meanwhile, we know there must be some point $p\in O$ at which $T^{ab}\neq\mathbf{0}$.  We can freely suppose that $t(p)\neq 0$ (because if $t(p)=0$, there necessarily exists a neighborhood around $p$ in which $T^{ab}\neq\mathbf{0}$, since $T^{ab}$ is smooth, and which must include points whose time values are greater and less than 0).  Suppose without loss of generality that $t(p)>0$ (if $t(p)<0$, simply reverse the temporal order of the ensuing argument---we have already chosen $O$ so that there are temporally prior, non-intersecting spacelike hypersurfaces).  Since $p\in O-\Sigma$, we know there's a spacelike hypersurface $\Sigma'$ that contains $p$ and slices $O$.  By the mass condition and the smoothness of $T^{ab}$, we know that $P^a(\Sigma')t_a>0$.  Now we can use Stokes' theorem as immediately above by connecting $\Sigma'$ and $\Sigma_+$ to reason to a contradiction.  Thus $\gamma$ cannot be spacelike.

\textbf{Case 3:} So far we have shown that if $\gamma$ is everywhere timelike then it must be (reparametrizable as) a geodesic, and that $\gamma$ cannot be everywhere spacelike.  The final case concerns curves that are sometimes timelike and sometimes spacelike.  Given case 1, it is sufficient to show that if $\gamma$ satisfies the assumptions of the theorem and is timelike at at least one point, then it is timelike everywhere.  Suppose otherwise---i.e., suppose there is at least one point $q$ at which $\gamma$ is spacelike.  Let $s_1\in I$ be such that $\gamma$ is timelike at $\gamma(s_1)$ and let $s_2\in I$ be such that $\gamma$ is spacelike at $\gamma(s_2)$.  Let $\xi^a$ be the tangent field to $\gamma$.  We can define a scalar field on $\gamma$ by $\alpha=\xi^at_a$.  $\alpha$ can be understood as a smooth function $\alpha:I\rightarrow\mathbb{R}$ defined by $\alpha(s)=\alpha\circ\gamma(s)=(\xi^a t_a)_{|\gamma(s)}$.  Since $\gamma$ is timelike at $\gamma(s_1)$, we know that $\alpha(s_1)>0$;  likewise, since $\gamma$ is spacelike at $\gamma(s_2)$, $\alpha(s_2)=0$.  Since $\alpha$ is just a smooth function on the reals, however, we know that there must be a number $t\in I$ such that $\alpha(t)>0$, but for which $\left(\frac{d}{ds}\alpha\right)(t)\neq0$.  But by definition of $\xi^a$, $\frac{d}{ds}\alpha(s)=(\xi^a{}_{|\gamma(s)})(\alpha)=\xi^a\nabla_a\alpha=t_b\xi^a\nabla_a\xi^b$.  So at $\gamma(t)$, we know that $(t_b\xi^a\nabla_a\xi^b)_{|\gamma(t)}\neq0$, and that $\xi^at_a>0$.

So $\gamma$ is timelike at $\gamma(t)$, which means (since $\gamma$ is smooth and imbedded) that there must be an open neighborhood $Q$ of $\gamma(t)$ such that the restriction of $\gamma[I]$ to $Q$ is timelike.  (Why?  Since $\gamma$ is smooth, there must be an open neighborhood $J\subseteq I$ of $t$ such that $\gamma[J]$ is timelike.  And since $\gamma$ is imbedded, there must be an open subset $Q$ of $M$ such that $\gamma[J]=\gamma[I]\cap Q$.  So the restriction of $\gamma[I]$ to $Q$ is timelike and contains $\gamma(t)$.)   We can freely choose $Q$ so that it is simply connected.  Note that since $\gamma$ is such that for any neighborhood of $\gamma$, there exists a smooth symmetric field $T^{ab}$ satisfying conditions \ref{mass}-\ref{non-vanishing2}, it follows that for any \emph{sub-}neighborhood $Q'$ of $Q$ containing $\gamma[I]\cap Q$, there also exists a smooth symmetric field $T^{ab}$ such that the restriction of $T^{ab}$ to $Q$ satisfies conditions \ref{mass}-\ref{non-vanishing2}, relative to $Q'$.  (Why? Extend $Q'$ to a neighborhood $O$ of all of $\gamma$ in any way at all, so long as $O\cap Q=Q'$.  Then a field $T^{ab}$ satisfying conditions \ref{mass}-\ref{non-vanishing2} relative to $O$ is guaranteed to exist by the assumptions of the theorem; the restriction of $T^{ab}$ to $Q$ automatically inherits conditions \ref{mass}-\ref{inside}.  And by the conservation of mass argument given in Lemma \ref{consMass}, if $T^{ab}$ is non-vanishing anywhere within $O$, as it must be, then it is possible to show by a series of flux integrals that it is non-vanishing along the length of the curve, and so $T^{ab}$ must be non-vanishing somewhere in $Q'$.)  But then if we take $Q$ as a submanifold of $M$ and take the restriction of $\gamma$ to $Q$ as a timelike curve, case 1 applies and $\gamma$ must be a geodesic everywhere in $Q$. It follows that at $\gamma(t)\in Q'$, $(\xi^a\nabla_a\xi^b)_{\gamma(t)}=0$, which is a contradiction (since we showed that $(t_b\xi^a\nabla_a\xi^b)_{|\gamma(t)}\neq0$).  And so $\gamma$ must be timelike everywhere.\hspace{.25in}$\square$

\section{Discussion}\label{sec-conclusion}

Mathematically, theorem \ref{W} differs from the Geroch-Jang theorem in at least two ways.  First, it requires a curvature condition: $R^{ab}{}_{cd}=\mathbf{0}$.  This condition enters the discussion via Lemma \ref{flatOp}, where the flat derivative operator used in the proof of Theorem \ref{W} is shown to exist.  Our method for constructing a flat derivative operator requires the existence of a rigid, non-rotating timelike field (a field $\eta^a$ such that $\nabla^{a}\eta^{b}=\mathbf{0}$).  The (local) existence of such a field in a spatially flat ($R^{abcd}=\mathbf{0}$) classical spacetime is in fact equivalent to $R^{ab}{}_{cd}=\mathbf{0}$.  Thus, without the curvature condition, our construction fails.  That said, it is quite likely (we believe) that a different argument can be given to show that an appropriate derivative operator does exist more generally, in which case it would be possible to relax the curvature condition in Theorem \ref{W}.

To evaluate whether this curvature condition is a defect of the present argument, however, one needs to consider the status of this condition in the context of the geometrized Newtonian gravitation. The condition is necessary to recover standard Newtonian gravitation from the geometrized theory (see appendix \ref{sec-N-C_theory}).  Without it, it is possible to find a more general ``Newtonian'' theory (see \citet{Kunzle, Ehlers, MalamentGR}), but with a vector potential replacing the scalar potential of standard Newtonian gravitation, and with a universal rotation field affecting the behavior of this vector potential.  We would like to note, however, that insofar as we were interested in the status of the geodesic principle in \emph{Newtonian} physics (rather than in some generalized Newtonian physics), $R^{ab}{}_{cd}=\mathbf{0}$ is a perfectly reasonable requirement: it holds just in case a geometrized Newtonian spacetime admits a standard Newtonian representation.  It is part of what makes a classical spacetime Newtonian.

The second difference is that the present result assumes the underlying manifold $M$ be simply connected; the Geroch-Jang theorem, however, does not seem to require any such global topological assumptions.  The reason that simple-connectedness is required here is that vector integration in a classical spacetime, at least as we have developed it, requires simple connectedness to ensure a unique result for the integral (since otherwise, parallel transport is not necessarily globally unique).  Geroch and Jang use Killing fields to avoid this problem entirely; however, in a classical spacetime one does not have access to timelike Killing fields, even locally or in flat spacetime.  However, there is a simple corollary available that (partially) extends the result to a more general case.
\begin{cor}\label{W1}
Let $(M,t_a,h^{ab},\nabla)$ be a classical spacetime, and suppose that $M$ is oriented.  Suppose also that $R^{abcd}=\mathbf{0}$ and $R^{ab}{}_{cd}=\mathbf{0}$.  For any $p\in M$, there exists a neighborhood of $p$, $Q$, such that if (1) $\gamma:I\rightarrow Q$ is a smooth curve, and (2) for any open subset $O$ of $Q$ containing $\gamma[I]$ there exists a smooth symmetric field $T^{ab}\in\mathfrak{T}^{\bullet}(M)$ satisfying conditions \ref{mass}-\ref{non-vanishing2} of Theorem \ref{W}, then $\gamma$ is a timelike curve that can be reparametrized as a geodesic (segment).
\end{cor}
Corollary \ref{W1} precisifies a sense in which \emph{local} geodesic motion has the status of a general theorem in geometrized Newtonian gravitation even in the absence of general topological assumptions.

\begin{acknowledgments}
I am indebted to David Malament for helpful comments on previous drafts of this paper, and for suggesting the topic.  Thank you, too, to helpful audiences in Paris and Wuppertal, and particularly to Harvey Brown and David Wallace.
\end{acknowledgments}

\appendix

\section{Review of Geometrized Newtonian Gravitation}\label{sec-N-C_theory}

In this appendix, we briefly review the central concepts of geometrized Newtonian gravitation.  We will not describe the full details of the theory; rather, the focus will be on setting up the language in which we operate in the body of the paper.  For details, we recommend \citet[Ch. 4]{MalamentGR}, which is (to our knowledge) the most systematic treatment of the subject available.

We begin by defining a classical spacetime.  \begin{defn}
A \emph{classical spacetime} is an ordered quadruple $(M, t_{ab}, h^{ab},\nabla)$, where  $M$ is a smooth, connected, four dimensional manifold; $t_{ab}$ is a smooth symmetric field on $M$ of signature $(1,0,0,0)$; $h^{ab}$ is a smooth symmetric field on $M$ of signature $(0,1,1,1)$; and $\nabla$ is a derivative operator on $M$ compatible with $t_{ab}$ and $h^{ab}$, i.e. it satisfies $\nabla_a t_{bc}=\nabla_a h^{bc}=\mathbf{0}$. We additionally require that $t_{ab}$ and $h^{ab}$ are orthogonal, i.e. $t_{ab}h^{bc}=\mathbf{0}$.\end{defn}

Note that ``signature,'' here, has been extended to cover the degenerate case.  We can see immediately from the signatures of $t_{ab}$ and $h^{ab}$ that neither is invertible.  Hence in general neither $t_{ab}$ nor $h^{ab}$ can be used to raise and lower indices.

The field $t_{ab}$ can be thought of as a temporal metric on $M$ in the sense that given any vector $\xi^a$ in the tangent space at a point, $p$, $||\xi^a||=(t_{ab}\xi^a\xi^b)^{1/2}$ is the \emph{temporal length} of $\xi^a$ at that point.  If the temporal length of $\xi^a$ is positive, $\xi^a$ is \emph{timelike}; otherwise, it is \emph{spacelike}.  At any point, it is possible to find a covector $t_a$, unique up to a sign, such that $t_{ab}=t_at_b$.  If there is a continuous, globally defined vector field $t_a$ such that at every point $t_{ab}=t_at_b$, then the spacetime is \emph{temporally orientable} (we encode the assumption that a spacetime is temporally oriented by replacing $t_{ab}$ with $t_a$ in our definitions of classical spacetimes).  $h^{ab}$, meanwhile, can be thought of as a spatial metric.  However, since there is no way to lower the indices of $h^{ab}$, we cannot calculate the spatial length of a vector directly.  Instead, we rely on the fact that if $\xi^a$ is a spacelike vector (as defined above), then there exists a (non-unique) covector $\sigma_a$ such that $\xi^a=h^{ab}\sigma_b$.  The \emph{spatial length} of $\xi^a$ can then be defined as $(h^{ab}\sigma_a\sigma_b)^{1/2}$.  It can be shown that this length is independent of the choice of $\sigma_a$.  If $\xi^a$ is not a spacelike vector, then there is no way to assign it a spatial length.  Note, too, that it is possible to define the Riemann curvature tensor $R^{a}_{\;\;bcd}$ and the Ricci tensor $R_{ab}$ with respect to $\nabla$ as in GR (or rather, as in differential geometry generally).  Flatness ($R^a_{\;\;bcd}=\mathbf{0}$) carries over intact from GR; we say a classical spacetime is \emph{spatially flat} if $R^{abcd}=R^a_{\;\;nmq}h^{bn}h^{cm}h^{dq}=\mathbf{0}$.  This latter condition is equivalent to $R^{ab}=h^{an}h^{bm}R_{nm}=\mathbf{0}$\cite{MalamentGR}.

We describe matter in close analogy with GR.  Massive point particles are represented by their worldlines, which are smooth future-directed timelike curves parameterized by elapsed time.  (Point particles in the current framework have the same attenuated status as in GR---really, we are thinking of a field theory, and point particles are some appropriate idealization.) For a point particle with mass $m$, we can always define a smooth unit vector field $\xi^a$ tangent to its worldline (the \emph{four-velocity}), such that we can define a \emph{four-momentum} field, $p^a=m\xi^a$.  Thus the mass of the particle is given by the temporal length of its four-momentum.  In similar analogy to the relativistic case, we can associate with any matter field a smooth symmetric field $T^{ab}$.  $T^{ab}$ encodes the four-momentum density of the matter field as determined by a future directed timelike observer at a point, but in this case all observers agree on the four-momentum density at any point $q$: $(p^a)_{|q}=(t_bT^{ab})_{|q}$.  Contracting once more with $t_b$ yields the mass density, $\rho=t_at_bT^{ab}$.  Since $T^{ab}$ encodes mass and momentum density in geometrized Newtonian gravitation, rather than energy and momentum density (as in GR), it is called the \emph{mass-momentum tensor}.  It is standard to assume that mass density is positive whenever $T^{ab}\neq\mathbf{0}$, i.e. $\rho=T^{ab}t_at_b>0$.  This condition, called the \emph{mass condition}, takes the place of the various energy conditions in GR.

In the present covariant four dimensional language, standard Newtonian mechanics can be expressed as follows.  Let $(M,t_a,h^{ab},\nabla)$ be a classical spacetime.  We require that $\nabla$ is flat.  We begin by considering the dynamics of a test point particle with mass $m$ and four-velocity $\xi^a$.  The acceleration of the particle's worldline, $\xi^b\nabla_b\xi^a$, is determined by the external forces acting on the particle according to the relation $F^a=m\xi^b\nabla_b\xi^a$.  In the absence of external forces, a massive test point particle undergoes geodesic motion.  If the total mass-momentum content of spacetime is described by $T^{ab}$, we require that the \emph{conservation condition} holds, i.e. at every point $\nabla_aT^{ab}=\mathbf{0}$.  To add gravitation to the theory, we can represent the gravitational potential as a smooth scalar field $\varphi$ on $M$.  $\varphi$ is required to satisfy Poisson's equation, $\nabla_a\nabla^a\varphi=4\pi\rho$ (where $\nabla^a$ is shorthand for $h^{ab}\nabla_b$).  Gravitation is considered a force; the gravitational force on a point particle is given by $F^a=-m\nabla^a\varphi$.

In geometrized Newtonian gravitation we again begin with a classical spacetime $(M,t_a,h^{ab},\nabla)$, but now we allow $\nabla$ to be curved.  Once again, the acceleration of a particle with mass $m$ and four-velocity $\xi^a$ is determined by the relation $F^a=m\xi^b\nabla_b\xi^a$, where $F^a$ represents the external forces acting on the particle; likewise, free massive test point particles undergo geodesic motion.  However, the geodesics are now determined relative to the not-necessarily-flat derivative operator.  The conservation condition is again expected to hold.  Gravitation enters the theory via a geometrized form of Poisson's equation: if $T^{ab}$ describes the total mass-momentum density in the spacetime, then the Ricci curvature tensor $R_{ab}=R^{n}_{\;\;abn}$ is given by $R_{ab}=4\pi\rho t_a t_b$.  Since the Riemann curvature tensor (and by extension, the Ricci tensor) is determined by $\nabla$, the geometrized Poisson's equation places a constraint on the derivative operator.  In particular, $\nabla$ must be such that, for all smooth vector fields $\xi^a$, $R_{ab}\xi^a=-2\nabla_{[b}\nabla_{n]}\xi^n=4\pi\rho t_a t_b\xi^a$.  Note, too, that the geometrized Poisson's equation forces spacetime to be spatially flat, because if Poisson's equation holds, then $R^{ab}=h^{an}h^{bm}R_{nm}=4\pi\rho h^{an}h^{bm}  t_n t_m=\mathbf{0}$ by the orthogonality condition on the metrics.

It is always possible to ``geometrize'' a gravitational field on a flat classical spacetime---that is, we can always move from the covariant formulation of standard Newtonian gravitation to geometrized Newtonian gravitation, via a result due to Andrzej Trautman.\cite{Trautman}
\begin{prop}[Trautman Geometrization Lemma.]\label{geometrization}
 \emph{(Slightly modified from \citet[Prop. 4.2.1.]{MalamentGR})} Let $(M,t_a,h^{ab},\overset{f}{\nabla})$ be a flat classical spacetime.  Let $\varphi$ and $\rho$ be smooth scalar fields on $M$ satisfying Poisson's equation, $\overset{f}{\nabla}_a\overset{f}{\nabla}\,^a\varphi=4\pi\rho$.  Finally, let $\overset{g}{\nabla}=(\overset{f}{\nabla},C^a_{\;\;bc})$,\footnote{This notation may require explanation. Briefly, if $\nabla$ is a derivative operator on $M$, then any other derivative operator on $M$ is determined relative to $\nabla$ by a smooth symmetric (in the lower indices) tensor field, $C^a_{\;\;bc}$, and so specifying the $C^{a}_{\;\;bc}$ field and $\nabla$ is sufficient to uniquely determine a new derivative operator.} with $C^a_{\;\;bc}=-t_bt_c\overset{f}{\nabla}\,^a\varphi$.  Then $(M,t_a,h^{ab},\overset{g}{\nabla})$ is a classical spacetime; $\overset{g}{\nabla}$ is the unique derivative operator on $M$ such that given any timelike curve with (normalized) tangent vector field $\xi^a$, \begin{equation}\label{geoEquiv}\tag{G}\xi^n\overset{g}{\nabla}_n\xi^a=\mathbf{0}\Leftrightarrow \xi^n\overset{f}{\nabla}_n\xi^a=-\overset{f}{\nabla}\,^a\varphi;\end{equation} and the Riemann curvature tensor relative to $\overset{g}{\nabla}$, $\overset{g}{R}\,^a_{\;\;bcd}$, satisfies \begin{align}&\overset{g}{R}_{ab}=4\pi\rho t_a t_b\tag{CC1}\label{CC1}\\ &\overset{g}{R}{}^a_{\;\;b}{}^c_{\;\;d}=\overset{g}{R}{}^{c}_{\;\;d}{}^a_{\;\;b}\tag{CC2}\label{CC2}\\
&\overset{g}{R}{}^{ab}{}_{cd}=\mathbf{0}\tag{CC3}\label{CC3}.\end{align}\end{prop}

Trautmann showed that it is also possible to go in the other direction.  That is, given a curved classical spacetime, it is possible to recover a flat classical spacetime and a gravitational field, $\varphi$---so long as the curvature conditions \eqref{CC1}-\eqref{CC3} are met.
\begin{prop}[Trautman Recovery Theorem.]\label{recovery}
 \emph{(Slightly modified from \citet[Prop. 4.2.5.]{MalamentGR})} Let $(M, t_a,h^{ab},\overset{g}{\nabla})$ be a classical spacetime that satisfies \eqref{CC1}-\eqref{CC3} for some smooth scalar field $\rho$.  Then, at least locally on $M$, there exists a smooth scalar field $\varphi$ and a flat derivative operator on $M$, $\overset{f}{\nabla}$, such that $(M,t_a,h^{ab},\overset{f}{\nabla})$ is a classical spacetime; \eqref{geoEquiv} holds for all timelike curves with (normalized) tangent vector field $\xi^a$; and $\varphi$ and $\overset{f}{\nabla}$ together satisfy Poisson's equation, $\overset{f}{\nabla}_a\overset{f}{\nabla}\,^a\varphi=4\pi\rho$.\end{prop} It is worth pointing out that the pair $(\overset{f}{\nabla},\varphi)$ is not unique.  It is also worth pointing out that whenever we begin with standard Newtonian theory and move to geometrized Newtonian theory, it is always possible to move back to the standard theory, because Prop. \ref{geometrization} guarantees that the curvature conditions \eqref{CC1}-\eqref{CC3} are satisfied.

\section{Integration in Classical Spacetimes}\label{sec-integration}

\subsection{Volume Elements and Hypersurfaces in Classical Spacetimes}

In what follows, we will make essential use of volume elements on differentiable manifolds with classical spacetime structure.  Some work is required to say what is meant by a volume element without a (invertible, non-degenerate) metric in the background.  First, the standard notion of orientability carries over intact from more familiar contexts: the underlying manifold of a classical spacetime is \emph{orientable} if it admits a smooth, globally defined, non-vanishing 4-form.  In this context, we can define a \emph{volume element} on an orientable manifold as a smooth 4-form $\epsilon_{abcd}$ satisfying the normalization condition,
\[\epsilon_{abcd}\epsilon_{efgh}h^{bf}h^{cg}h^{dh}=6t_at_e,\] which is equivalent to requiring that, given any four vectors at any point $p\in M$, if one of them is a unit timelike vector, $\xi^a$, and the other three are mutually orthogonal unit spacelike vectors, $\overset{i}{\eta}{}^a$, then $\epsilon_{abcd}\xi^a\overset{1}{\eta}{}^a\overset{2}{\eta}{}^a\overset{3}{\eta}{}^a=\pm 1$.  Dimensionality considerations are sufficient to show that the volume element is unique up to sign.  Specifying a volume element on $M$ provides an orientation for the manifold; when we call a manifold \emph{oriented}, we are assuming a fixed choice of a volume element in the background.  Finally, to say two n-forms $\omega_{a_1\cdots a_n}$ and $\omega'_{a_1\cdots a_n}$ are \emph{co-oriented} is to say that $\omega_{a_1\cdots a_n}=f\omega'_{a_1\cdots a_n}$, where $f>0$ everywhere.

A hypersurface in a classical spacetime is \emph{spacelike} at a point if all of its tangent vectors are; otherwise it is \emph{timelike} at that point.  In what follows, we will limit attention to hypersurfaces that are either everywhere spacelike or everywhere timelike.  Suppose $\Sigma$ is a (timelike or spacelike) hypersurface of $M$.  As above, we will say $\Sigma$ is orientable if it admits a smooth, globally defined, non-vanishing 3-form. Then, if $\Sigma$ is orientable, it is always possible to factor the volume element on $M$ in the neighborhood of $\Sigma$ into $\overset{M}{\epsilon}_{abcd}=\overset{\Sigma}{n}\,_{[a}\overset{\Sigma}{\omega}_{bcd]}$, where $\overset{\Sigma}{\omega}_{abc}$ is a (non-unique) 3-form on $M$ and where $\overset{\Sigma}{n}_a$ is a unit covector field normal to $\Sigma$. If $\Sigma$ is spacelike, then $\overset{\Sigma}{n}_a=\pm t_a$; if $\Sigma$ is timelike, then $h^{ab}\overset{\Sigma}{n}_a\overset{\Sigma}{n}_b=1$ and whenever $v^a\in\mathfrak{T}^{\bullet}(M)$ is tangent to $\Sigma$, $v^a\overset{\Sigma}{n}_a=0$.  We can then take $\overset{\Sigma}{\imath}\,^*(\overset{\Sigma}{\omega}_{abc})=\overset{\Sigma}{\epsilon}_{abc}$ to define a volume element on $\Sigma$ (in other words, the restriction  to $\Sigma$ of any 3-form satisfying the factorization condition above gives a volume element on $\Sigma$).  As above, dimensionality considerations show that volume elements on hypersurfaces are unique up to sign; to say a hypersurface is oriented will be to assume that there's a fixed choice of volume element in the background.

Note that there are in general two possible unit covector fields normal to any given oriented hypersurface of $M$: if $\overset{\Sigma}{n}_a$ is a unit normal covector field, then so is $-\overset{\Sigma}{n}_a$.  However, the sign of $\overset{\Sigma}{n}_a$ as we have defined it is wholly fixed by the relative orientations of $M$ and $\Sigma$ because $\overset{M}{\epsilon}_{abcd}$ is fixed by the orientation of $M$ and the sign of $\overset{\Sigma}{\omega}_{abc}$ is fixed by the orientation of $\Sigma$.  Thus given any oriented hypersurface of $M$, there is a unique unit normal covector field that satisfies the stated factorization condition.  Conversely, a choice of normal covector field uniquely picks out an orientation for a hypersurface.  As a matter of definition, in the special case where $\Sigma$ is an oriented \emph{spacelike} hypersurface, we will call $\Sigma$ \emph{future-directed} (relative to the orientation of $M$) if $\overset{\Sigma}{n}_a=t_a$; likewise, $\Sigma$ is \emph{past-directed} if $\overset{\Sigma}{n}_a=-t_a$.  Finally, if $A$ is an oriented $p$ dimensional manifold, we will denote its volume element by $\overset{A}{\epsilon}_{a_1\cdots a_p}$.

\subsection{Integration in Flat Classical Spacetimes}

Here we assume that $\nabla$ is a flat derivative operator and that $M$ is oriented and simply connected.  In the body of the paper, we need to make sense of some improper-looking integrals, in which the integrand and the integral have (the same) contravariant indices.  That is, we will consider integrals of the form $\alpha^{a_1\cdots a_n}=\int_{S}\beta^{a_1\cdots a_n}\omega_{b_1\cdots b_{p}}$ where $S$ is a three or four dimensional imbedded submanifold of $M$ and $\omega$ is a $3-$ or $4-$form, respectively.  We make no claims about what such integrals mean (if anything) under general circumstances.  However, when $\nabla$ is flat and $M$ is orientable and simply connected, they can be understood as follows.  Pick a point, $q\in M$, and let $\{\overset{1}{\sigma}_a(q),\ldots,\overset{4}{\sigma}_a(q)\}$ be an orthonormal$^*$ (the star indicates that the language is being abused) basis for the cotangent space of $M$ at $q$.   Since $\nabla$ is flat, parallel transport of covectors is (locally) path-independent; since $M$ is simply connected, we can extend the cobasis at $q$ to all points in $M$ without introducing any ambiguities, by parallel transporting each of the cobasis elements to each other point.  This method is guaranteed to produce smooth fields of orthonormal covectors on $M$---that is, fields of constant basis covectors, $\{\overset{1}{\sigma}_a,\ldots,\overset{4}{\sigma}_a\}$.

We can define the integrals required in terms of such bases.  Taking an integral with a single contravariant index (it is easy to see how to generalize to more indices), we say $\alpha^{a}=\int_{S}\beta^{a}\omega_{b_1\cdots b_p}$ is the vector field such that, given any covector field $\kappa_a\in\mathfrak{T}_{\bullet}(S)$, $\alpha^a\kappa_a=\sum_{i=1}^4\overset{i}{\kappa}\,\overset{i}{\sigma}_a\alpha^a=\sum_{i=1}^4\overset{i}{\kappa}\,\int_{S}\overset{i}{\sigma}_a\beta^{a}\omega_{b_1\cdots b_p}$, where $\overset{i}{\kappa}$ is defined so that $\kappa_a=\sum_{i=1}^{4}\overset{i}{\kappa}\overset{i}{\sigma}_a$.  Note that since $S$ is an imbedded submanifold of $M$, $\overset{S}{\imath}\,^*(\beta^a\overset{i}{\sigma}_a)=\beta^a\overset{i}{\sigma}_a\circ \imath=\beta^a\overset{i}{\sigma}_a$ because $\beta^a\overset{i}{\sigma}_a$ is a scalar field.  The vector $\alpha$ must exist, as the defining relation for the integral generates a map from the covectors to $C^{\infty}$.  Moreover, it can easily be shown that this definition of the integral is independent of the choice of basis, due to the linearity of the integral.

Finally, it will prove helpful to register up front how to express two well-known facts about integration in the present language.  First, suppose that $\Sigma\subset M$ is an oriented, imbedded hypersurface of $M$ and let $\beta^a$ be an arbitrary contravariant vector field on $M$.  Then we can immediately write $4\beta^a\overset{M}{\epsilon}_{abcd}=4\beta^a\overset{\Sigma}{n}_{[a}\overset{\Sigma}{\omega}_{bcd]}= \beta^a\overset{\Sigma}{n}_a\overset{\Sigma}{\omega}_{bcd}-3\overset{\Sigma}n_{[b}\beta^a\overset{\Sigma}{\omega}_{|a|cd]}$.  To integrate, we need to take the pull-back to $\Sigma$ of both sides of this expression, yielding $\overset{\Sigma}{\imath}\,^*(4\beta^a\overset{M}{\epsilon}_{abcd}) =\overset{\Sigma}{\imath}\,^*(\beta^a\overset{\Sigma}{n}_a\overset{\Sigma}{\omega}_{bcd}-3\overset{\Sigma}n_{[b}\beta^a\overset{\Sigma}{\omega}_{|a|cd]}) =\overset{\Sigma}{\imath}\,^*(\beta^a\overset{\Sigma}{n}_a\overset{\Sigma}{\omega}_{bcd})= \overset{\Sigma}{\imath}\,^*(\beta^a\overset{\Sigma}{n}_a)\overset{\Sigma}{\epsilon}_{bcd}$, because the pull-back map commutes with exterior multiplication, and $\overset{\Sigma}{\imath}\,^*(\overset{\Sigma}{n}_a)=\mathbf{0}$ because $\overset{\Sigma}{n}_a$ is normal to $\Sigma$.  Thus,
\[\int_{\Sigma}\overset{\Sigma}{\imath}\,^*(\beta^{a}\overset{M}{\epsilon}_{a bcd})=\frac{1}{4}\int_{\Sigma}\overset{\Sigma}{\imath}\,^*(\beta^{a}\overset{\Sigma}{n}_a)\overset{\Sigma}{\epsilon}_{bcd}.\]   Secondly, suppose that $N$ is a four dimensional submanifold of $M$ with boundary $\partial N$, where we assume $\partial N$ can be written as the union of a collection of hypersurfaces, each of which is everywhere timelike or everywhere spacelike.  Then if $\omega_{bcd}$ is any $3-$form on $N$, we can write Stokes' theorem in the current language as
\[\int_Nd_a\omega_{bcd}=\int_N\nabla_{[a}\omega_{bcd]}=\int_{\partial N}\overset{\partial N}{\imath}{}^*(\omega_{bcd}),\] where $d$ represents the exterior derivative on $N$.

\section{Supplementary proofs}\label{sec-proofs}

\textbf{Proof of Prop. \ref{consMom}.}  Let $\Sigma_1$ and $\Sigma_2$ be two future-directed spacelike hypersurfaces slicing the support of $T^{ab}$.  Consider a third (timelike) hypersurface, $\Sigma_3$, connecting $\Sigma_1$ and $\Sigma_2$ in such a way that (1) $\supp{T^{ab}}\cap\Sigma_3=\emptyset$ and (2) if we reverse the orientation of the temporally prior of the spacelike hypersurfaces (say, $\Sigma_2$), then $\partial S\equiv\Sigma_1\cup\Sigma_2^{-}\cup\Sigma_3$ forms the (outwardly oriented) boundary of an oriented, simply connected four dimensional submanifold $S$ of $M$.  Since the support of $T^{ab}$ does not intersect $\Sigma_3$, it follows immediately that $\int_{\Sigma_3}T^{ab}\overset{\Sigma_3}{n}_b\overset{\Sigma_3}{\epsilon}_{cde}=\mathbf{0}$.  Let $\kappa_a$ be an arbitrary covector field on $M$.  Then by Stokes' theorem and the relation above concerning flux integrals,
\begin{align*}
\kappa_a(P^a&(\Sigma_1)-P^a(\Sigma_2))\\
&=\sum_{i=1}^4\overset{i}{\kappa}\left(\int_{\Sigma_1}T^{ab}\overset{i}{\sigma}_at_b\overset{\Sigma_1}{\epsilon}_{cde}- \int_{\Sigma_2}T^{ab}\overset{i}{\sigma}_at_b\overset{\Sigma_2}{\epsilon}_{cde}\right)\\
&=4\sum_{i=1}^4\overset{i}{\kappa}\left(\int_{S}\nabla_{[n}T^{ab}\overset{i}{\sigma}_{|a}\overset{S}{\epsilon}_{b|cde]}\right)
\end{align*}
The third equality follows because $T^{ab}t_a\overset{i}{\sigma}_b$ is a scalar field, and so it is unaffected by the pull-backs; the fifth equality makes use of the relation cited above concerning flux integrals; and the final equality follows by Stokes' theorem.

Consider the integrand of the last of the expressions above, $\nabla_{[n}T^{ab}\overset{i}{\sigma}_{|a}\overset{S}{\epsilon}_{b|cde]}$.  The space of $n-$forms on any $n$ dimensional manifold is one dimensional, and so it must be that $\nabla_{[n}T^{ab}\overset{i}{\sigma}_{|a}\overset{S}{\epsilon}_{b|cde]}=f\; \overset{S}{\epsilon}_{ncde}$, for some scalar field $f$.  The goal is to show that $f$ must be zero; if this is the case, then the integrand vanishes.  Let $\overset{S}{\epsilon}\,^{abcd}$ (with raised indices) be a totally anti-symmetric contravariant tensor, normalized so that $\overset{S}{\epsilon}_{abcd}\overset{S}{\epsilon}\,^{efgh}=4!\delta_a{}^{[e}\delta_b{}^f\delta_c{}^g\delta_d{}^{h]}$.  This field can be constructed out of any (contravariant) basis fields for $S$.  Multiplying the integrand by $\overset{S}{\epsilon}\,^{abcd}$ and contracting, then, we find
\begin{align*}
f\overset{S}{\epsilon}_{ncde}\overset{S}{\epsilon}\,^{ncde}=4!f&=\nabla_{[n}(T^{ab}\overset{i}{\sigma}_{|a}\overset{S}{\epsilon}_{b|cde]})\overset{S}{\epsilon}\,^{ncde}\\
&=4!\nabla_n(T^{ab}\overset{i}\sigma_a)\delta_b{}^n=4!\nabla_bT^{ab}\overset{i}{\sigma}_a=0,
\end{align*}
where the last step follows from the conservation condition on $T^{ab}$.  Thus $f=0$.  It follows immediately that $\kappa_a(P^a(\Sigma_1)-P^a(\Sigma_2))=0$.  But $\kappa_a$ was an arbitrary covector, which means that $P^a(\Sigma_1)-P^a(\Sigma_2)$ must vanish identically, and so $P^a(\Sigma_1)=P^a(\Sigma_2)$.
\hspace{.25in}$\square$

\textbf{Proof of Prop. \ref{delAngMom}.}  Fix $o\in M$ and consider any $p\in M$ and any spacelike hypersurface $\Sigma$ that slices the support of $T^{ab}$.  Then $(J^{ab})_{|p}=\int_{\Sigma}\overset{p}{\chi}\,^{[a}T^{b]c}t_c\overset{\Sigma}{\epsilon}_{def} = \int_{\Sigma}\overset{o}{\chi}\,^{[a}T^{b]c}t_c\overset{\Sigma}{\epsilon}_{def} + \int_{\Sigma}(\overset{p}{\chi}\,^{[a}-\overset{o}{\chi}\,^{[a})T^{b]c}t_c\overset{\Sigma}{\epsilon}_{def}$, where in the last step we have added and subtracted $\int_{\Sigma}\overset{o}{\chi}\,^{[a}T^{b]c}t_c\overset{\Sigma}{\epsilon}_{def}$, which is a vector that we can understand to be defined at $p$.  Notice that $(\overset{p}{\chi}\,^{a}-\overset{o}{\chi}\,^{a})$ is a constant vector field: at any point $q$, it is just the vector ``from $p$ to $q$'' minus the vector ``from $o$ to $q$''.  Thus the field $(\overset{p}{\chi}\,^{a}-\overset{o}{\chi}\,^{a})$ is given by the constant vector ``from $p$ to $o$'' at every point.  This could be characterized as $(\overset{p}{\chi}\,^a)_{|o}$ parallel transported to every point or alternatively as $-(\overset{o}{\chi}\,^a)_{|p}$ parallel transported to every point.  For clarity, we will use the notation $(v^a)_{\parallel p}$ to represent the (global) vector field found by parallel transporting $(v^a)_{|p}$ to all points.  In this notation, we have $(J^{ab})_{|p}=\int_{\Sigma}\overset{o}{\chi}\,^{[a}T^{b]c}t_c\overset{\Sigma}{\epsilon}_{def} - \int_{\Sigma}(\overset{o}{\chi}\,^{[a})_{\parallel p}T^{b]c}t_c\overset{\Sigma}{\epsilon}_{def}$.

Since $(\overset{o}{\chi}\,^{a})_{\parallel p}$ is a constant vector field, we can pull it out of the integral to write, $(J^{ab})_{|p}=\int_{\Sigma}\overset{o}{\chi}\,^{[a}T^{b]c}t_c\overset{\Sigma}{\epsilon}_{def} -\left(\overset{o}{\chi}_{\parallel p}\,^{[a}\int_{\Sigma}T^{b]c}t_c\overset{\Sigma}{\epsilon}_{def}\right)_{|p}$.  But $\left((\overset{o}{\chi}\,^a)_{\parallel p}\right)_{|p}=(\overset{o}{\chi}\,^a)_{|p}$ and $\int_{\Sigma}T^{bc}t_c\overset{\Sigma}{\epsilon}_{def}=P^b$, so we have $(J^{ab})_{|p}= \int_{\Sigma}\overset{o}{\chi}\,^{[a}T^{b]c}t_c\overset{\Sigma}{\epsilon}_{def}- (\overset{o}{\chi}\,^{[a}P^{b]})_{p}$.  Moreover, in the present notation, $\int_{\Sigma}\overset{o}{\chi}\,^{[a}T^{b]c}t_c\overset{\Sigma}{\epsilon}_{def}=(J^{ab})_{\parallel o}$.  This means we can write $(J^{ab})_{|p}=\left((J^{ab})_{\parallel o}-\overset{o}{\chi}\,^{[a}P^{b]}\right)_{|p}$.  But $p$ was arbitrary, so $J^{ab}$ can be characterized in general as $J^{ab}=(J^{ab})_{\parallel o}-\overset{o}{\chi}\,^{[a}P^{b]}$.  Taking the action of $\nabla_a$ on both sides of this final expression yields $\nabla_a J^{bc}=-\delta_a{}^{[b}P^{c]}$. \hspace{.25in}$\square$

\textbf{Proof of Prop. \ref{uniqueCoM}.}  First we will prove that a point as described in the statement of the proposition exists.  Fix some arbitrary $o\in \Sigma$ and consider $(J^{ab}t_b)_{|o}/(P^nt_n)=\int_\Sigma \overset{o}{\chi}\,^aT^{bc}t_bt_c\overset{\Sigma}{\epsilon}_{def}/(P^nt_n)=R^a$. Note that this expression is simply a definition of $R^a$---no claim has yet been made; moreover, $P^nt_n$ is just a scalar constant.  We have used the fact that since $o\in\Sigma$, $\overset{o}{\chi}\,^a$ is spacelike on all of $\Sigma$ to simplify this expression.  $R^a$ is a constant, spacelike vector field (spacelike because the integrand is spacelike over the entire domain of integration).  We can then write $\int _{\Sigma} \overset{o}{\chi}\,^aT^{bc}t_bt_c\overset{\Sigma}{\epsilon}_{def}=R^a\int_{\Sigma}T^{bc}t_bt_c \overset{\Sigma}{\epsilon}_{def}=\int_{\Sigma} R^aT^{bc}t_bt_c \overset{\Sigma}{\epsilon}_{def}$ or $\int_{\Sigma}(\overset{o}{\chi}\,^a-R^a)T^{bc}t_bt_c \overset{\Sigma}{\epsilon}_{def}=\mathbf{0}$.  But $\Sigma$ is a  spacelike hypersurface of a geodesically complete, simply connected classical spacetime, so it is a flat, three dimensional Euclidean manifold.  Thus $\overset{o}{\chi}\,^a-R^a$ would be the position vector field centered at the point $q=o+R^a(o)$ (where we are using the natural affine structure of Euclidean space to represent points as a formal sum between a point and a vector, so a point $p$ can be written as a sum of any point $p'$ and a vector $v$ from $p'$ to $p$ as $p=p'+v$), if in fact there is such a point in $\Sigma$.  But even if there is no such $q$ in $\Sigma$, the vector field $\overset{o}{\chi}\,^a-R^a$ is well defined, and we can use the notation $\overset{o}{\chi}\,^a-R^a=\overset{q}{\chi}\,^a$ to describe a vector field on $\Sigma$ without assuming that $q\in\Sigma$.  Note, however, that if $q\in\Sigma$, then $(J^{ab}t_b)_{|q}=\mathbf{0}$ and $q$ would be the desired point, so it only remains to show that $q\in\Sigma$ and we will have established existence.

We claim that there is such a point $q\in\Sigma$.  To see why, first note that $\int_{\Sigma}\overset{q}{\chi}\,^aT^{bc}t_bt_c \overset{\Sigma}{\epsilon}_{def}$ is a positively weighted average of position vectors, and so it can only vanish if the position origin falls within the spacelike slice of the convex hull of $T^{ab}$ over which the average is performed.  (See, for instance, \citet{Benson} for a proof of this well-known claim.)  So $q\in \CH{T^{ab}}$ (and \emph{a fortiori}, $q\in M$, since $M$ is geodesically complete).  But $\Sigma$ slices the spatial convex hull of $T^{ab}$, by hypothesis.  So suppose there is no such $q$ in $\Sigma$.  Then we could define $\tilde{\Sigma}=\Sigma\cup\{q\}$.  Since $q$ is spacelike related to $o\in\Sigma$, $\tilde{\Sigma}$ is a spacelike hypersurface.  Thus we have a spacelike hypersurface such that $\Sigma\subseteq\tilde{\Sigma}$ but $\Sigma\cap\CH{T^{ab}}\neq\tilde{\Sigma}\cap\CH{T^{ab}}$, and so $\Sigma$ does not slice $\CH{T^{ab}}$, which is a contradiction.  Thus, since $q\in\CH{T^{ab}}$ and $q$ is spacelike related to $o\in\Sigma$ (as it is by construction), $q\in \Sigma$.

It remains to show that $q$ is unique.  Suppose there were two such points, $q$ and $q'$, where $q\neq q'$.  Then $\int_{\Sigma} \overset{q}{\chi}\,^aT^{bc}t_bt_c \overset{\Sigma}{\epsilon}_{def}=\int_{\Sigma} \overset{q'}{\chi}\,^aT^{bc}t_bt_c \overset{\Sigma}{\epsilon}_{def}=\mathbf{0}=\int_{\Sigma} (\overset{q}{\chi}\,^a-\overset{q'}{\chi}\,^a)T^{bc}t_bt_c \overset{\Sigma}{\epsilon}_{def}$.  Let $R^a$ be as defined above and furthermore take $Q^a$ be the unique constant vector field such that $q'=o+Q^a(o)$.  Then we have $\int_{\Sigma} (\overset{q}{\chi}\,^a-\overset{\;q'}{\chi}\,^a)T^{bc}t_bt_c \overset{\Sigma}{\epsilon}_{def}=\int_{\Sigma}(R^a-Q^a)T^{bc}t_bt_c \overset{\Sigma}{\epsilon}_{def}=(R^a-Q^a)\int_{\Sigma}T^{bc}t_bt_c \overset{\Sigma}{\epsilon}_{def}=\mathbf{0}$.  But $T^{bc}t_bt_c$ is nonvanishing and never negative by assumption (the first follows because $T^{ab}$ is nonvanishing and the second by the mass condition), and so $\int_{\Sigma}T^{bc}t_bt_c \overset{\Sigma}{\epsilon}_{def}\neq 0$.  Thus $R^a-Q^a=\mathbf{0}$ and $q=q'$.  It follows that $q$ is unique.\hspace{.25in}$\square$

\textbf{Proof of Lemma \ref{flatOp}.} All of the propositions of the form X.X.X cited in this proof are references to \citet{MalamentGR}; we will refer to the proposition numbers directly and suppress further citations where no ambiguity can arise.

There are many flat derivative operators compatible with $h^{ab}$ and $t_a$ (see Prop. 4.2.5). Our strategy will be to start with one such operator and then use it construct a second operator that additionally satisfies the second condition of the proposition.

Since $R^{abcd}=\mathbf{0}$ and $R^{ab}{}_{cd}=\mathbf{0}$, there exists (globally, since $M$ is simply connected) a timelike vector field $\eta^a$ that is rigid and non-rotating (i.e. $\nabla^a\xi^b=\mathbf{0}$).  Let $\hat{h}_{ab}$ be the spatial projection field relative to $\eta^a$ (see Prop. 4.1.2) and define $\phi^a=\eta^n\nabla_n \eta^a$ and $\kappa_{ab}=\hat{h}_{n[b}\nabla_{a]}\eta^n$.  We will take the reference derivative operator to be given by $\overset{f1}{\nabla}=(\nabla,\overset{01}{C^{a}}_{bc})$ where $\overset{01}{C^{a}}_{bc}=2h^{am}t_{(b}\kappa_{c)m}$.  As is shown in the proof of Prop. 4.2.5, this choice of derivative operator is flat and compatible with $t_a$ and $h^{ab}$.

Prop. 4.2.5 shows that a second derivative operator/vector field pair $(\overset{f2}{\nabla},\overset{2}{\phi}{}^a)$ will also be flat and compatible with $h^{ab}$ and $t_a$ iff $\nabla^a(\overset{2}{\phi}{}^b-\overset{1}{\phi}{}^b)=0$ and $\overset{f2}{\nabla}=(\overset{f1}{\nabla},\overset{12}{C^a}_{bc})$ where $\overset{12}{C^a}_{bc}=t_bt_c(\overset{2}{\phi}{}^a-\overset{1}{\phi}{}^a)$.  Moreover, by Prop. 1.7.3, there must exist a symmetric tensor field $\overset{02}{C^a}_{bc}$ such that $\overset{f2}{\nabla}=(\nabla,\overset{02}{C^a}_{bc})$.  Indeed, $\overset{02}{C^a}_{bc}=\overset{01}{C^a}_{bc}+\overset{12}{C^a}_{bc}$.

One can write the required relation between $\overset{1}{\phi}{}^a$ and $\overset{2}{\phi}{}^a$ as $\overset{2}{\phi}{}^a=\overset{1}{\phi}{}^a+\psi^a$ where $\psi^a$ is a covariant spacelike vector field satisfying $\nabla^b\psi^a=0$.  The condition that two derivative operators agree at a point $p$ can be stated by demanding that the $C^{a}_{\;\;\;bc}$ field relating them vanishes at that point.  Thus $\overset{f2}{\nabla}$ agrees with $\nabla$ on $\gamma$ just in case $\overset{02}{C^a}_{bc}$ vanishes on $\gamma$.  This condition in turn holds just in case $\overset{01}{C^a}_{bc}+\overset{12}{C^a}_{bc}=2h^{am}t_{(b}\kappa_{c)m}+t_bt_c\psi^a=0$ on $\gamma$.  Since $\eta^a$ is timelike, $2t_{(b}\kappa_{c)}^{\;\;a}+t_bt_c\psi^a=0$ on $\gamma$ just in case $\eta^b\eta^c(2t_{(b}\kappa_{c)}^{\;\;a}+t_bt_c\psi^a)=0$ on $\gamma$.  But $\eta^bt_b=\eta^ct_c=1$ and, as shown in the proof of Prop. 4.2.5, $2\kappa_{a}^{\;\;b}\eta^a=\overset{1}{\phi}{}^b$.  Thus $\eta^b\eta^c(2t_{(b}\kappa_{c)}^{\;\;a}+t_bt_c\psi^a)=\overset{1}{\phi^a}+\psi^a$, and so $\overset{f2}{\nabla}$ agrees with $\nabla$ on $\gamma$ whenever $\psi^a=-\overset{1}{\phi}{}^a$ on $\gamma$.  Note that this condition is equivalent to saying that, again on $\gamma$, $\overset{2}{\phi}{}^{a}=0$.

As stated above, it is also necessary that $\nabla^b\psi^a=0$ obtain.  So we have two conditions on $\psi^a$ (that it is constant in spacelike directions, and that it is the opposite of $\overset{1}{\phi}{}^a$ on $\gamma$).  We claim that there is a field that meets both conditions.  For any spacelike hypersurface $\Sigma$ slicing the spatial convex hull of $T^{ab}$, let $\psi^a$ be the vector field one finds by parallel transporting (relative to $\nabla$) the vector $-\overset{1}{\phi}{}^a$ at the point where $\gamma$ intersects $\Sigma$ to all other points of $\Sigma$ (this construction cannot produce ambiguities because we have assumed spatial flatness, and thus parallel transport in space is always path-independent, at least in a simply connected manifold).  Then $\psi^a$ is smooth, because $\overset{1}{\phi}{}^a$ is, and moreover, it satisfies both requirements.  Thus $\overset{f2}{\nabla}=(\overset{f1}{\nabla},t_bt_c\psi^a)=(\nabla,2h^{am}t_{(b}\kappa_{c)m}+t_bt_c\psi^a)$ is the required derivative operator.\hspace{.25in}$\square$

\textbf{Proof of Lemma \ref{consMass}.}  This result follows the proof of Prop. \ref{consMom} closely.  The most important thing to note is that here we assume that $\nabla_aT^{ab}=\mathbf{0}$, but not that $\overset{f}{\nabla}_aT^{ab}=\mathbf{0}$.  Thus the argument that the integrand $\overset{f}{\nabla}_{[n}T^{ab}t_{|a}\overset{S}{\epsilon}_{b|cde]}$ vanishes fails. However, we now are considering a special case wherein $\kappa_a=t_a$.  Without loss of generality, we can always choose to integrate relative to a set of basis fields in which $t_a$ is a basis element.  Then, by the Stokes' theorem argument given in the proof of Prop. \ref{consMom}, we have $t_a(P^a(\Sigma_1)-P^a(\Sigma_2))=\int_{S}\overset{f}{\nabla}_{[n}T^{ab}t_{|a}\overset{S}{\epsilon}_{b|cde]}$.  But $\overset{f}{\nabla}_{[n}T^{ab}t_{|a}\overset{S}{\epsilon}_{b|cde]}$ is an exterior derivative, and so it is invariant under different choices of covariant derivative operator.  That is, we can write $\overset{f}{\nabla}_{[n}T^{ab}t_{|a}\overset{S}{\epsilon}_{b|cde]}=d_n(T^{ab}t_{a}\overset{S}{\epsilon}_{bcde})=\nabla_{[n}T^{ab}t_{|a}\overset{S}{\epsilon}_{b|cde]}$, where in the last expression we are using the general curved derivative operator associated with the spacetime---relative to which $T^{ab}$ \emph{is} conserved.  Again by reasoning present in the proof to Prop. \ref{consMom}, it can be shown that $\nabla_{[n}T^{ab}t_{|a}\overset{S}{\epsilon}_{b|cde]}=\nabla_b(T^{ab}t_a)\overset{S}{\epsilon}_{ncde}$.  Since $t_a$ is compatible with $\nabla$, we have $\nabla_b(T^{ab}t_a)=0$.  Thus $t_a(P^a(\Sigma_1)-P^a(\Sigma_2))=0$, or for any spacelike hypersurfaces slicing the support of $T^{ab}$, $\Sigma_1$ and $\Sigma_2$, $P^a(\Sigma_1)t_a=P^a(\Sigma_2)t_a$.  \hspace{.25in}$\square$

\end{document}